\documentclass[12pt]{article}
\usepackage{epsfig}
\usepackage{amsmath}
\usepackage{amssymb}
\usepackage{amsfonts}
\usepackage{array}
\usepackage{theorem}

\theoremstyle{break}
\theoremheaderfont{\normalfont\bfseries}
\def\super       {\hat{sl}(2|1;\mathbb C)_k}
\def\su          {\hat{su}(2)} 
\def\iso         {\hat{sl}(2;\mathbb C)}

\def\defd        {\stackrel{\rm d}{=}}
\def\half        {\textstyle{1\over2}}
\def\quarter     {\textstyle{1\over4}}
\def\Zplus       {\mathbb Z _{+}}
\def\N           {\mathbb N}

\def\Z           {\mathbb Z}
\def\R           {\mathbb R}

\def\ben         {\begin{equation}}
\def\een         {\end{equation}}
\def\bea         {\begin{eqnarray}}
\def\eea         {\end{eqnarray}}

\def\im          {\mathop{\mathrm{Im}}}
\def\hslc        {\hat{sl}(2|1;{\mathbb C})}
\def\hslck       {\hat{sl}(2|1;{\mathbb C})_k}

\def\hf          {\frac{1}{2}}

\begin{document}
\def\theequation{\thesection.\arabic{equation}}
\begin{titlepage}
\begin{flushright}
DTP/98/7\\
March 1998\\
\end{flushright}
\vspace{1cm}
\begin{center}
{\Large\bf Admissible $\super$ Characters and Parafermions.}\\
\vspace{1cm}
{\large M. ~Hayes and 
A. ~Taormina}\\
\vspace{0.5cm}
{\it Department of Mathematical Sciences, University of Durham, Durham, DH1 ~3LE, England}\\
\vspace{2cm}
\end{center}
%
%
%
%
\begin{abstract}
The branching functions of the affine superalgebra $\hslck$ characters into characters of the subalgebra $\iso_k$ are calculated for fractional levels
$k=\frac{1}{u}-1, ~u \in \N$. They involve rational torus $A_{u(u-1)}$ and $\Z_{u-1}$ parafermion characters.
\end{abstract}
\vskip 10truecm
{}e-mail: M.R.Hayes@durham.ac.uk, Anne.Taormina@durham.ac.uk
\end{titlepage}
\section{Introduction.}
\renewcommand{\theequation}{1.\arabic{equation}}
\setcounter{equation}{0}

It has long been stressed that non-critical strings might well be described
in terms of a topological $G/G$ Wess-Zumino-Novikov-Witten model \cite{Y,FY,HY},
where $G$ is a Lie supergroup or a Lie group, depending on whether the string
theory considered is supersymmetric or not. The exact correspondence between
the traditional approach to non-critical strings and the latter is yet
to be proven. However, a crucial ingredient in the
description of the spectrum in the $G/G$ picture is the representation theory 
of the corresponding
affine Lie (super)algebra, $\hat{g}$, at fractional level $k= p/u-h,~~~
p\in \Z \setminus \{0\}, u \in \N,~~~{\rm gcd}(p,u)=1$, with $h$ the dual Coxeter number of $\hat{g}$.

In this paper we investigate further the characters of the complex affine
superalgebra $\super$ because of its potential relevance to the description
of $N=2$ non-critical strings. Indeed, when  the matter, which is coupled to supergravity in the $N=2$ non-critical string,
is minimal, i.e. taken in an $N=2$ super Coulomb gas representation with
central charge,
\ben
c_{\text{matter}}=3\left(1-\frac{2p}{u}\right),~~~~p, u \in \N,~~~{\rm gcd}(p,u)=1,
\end{equation}
the level of the `matter' affine superalgebra $\hslc$ appearing in the 
\newline $SL(2|1;\R)/SL(2|1;\R)$ model is of the form
\ben
k = \frac{p}{u} - 1.
\label{levelp}
\end{equation}
That is, the level precisely takes values for which admissible representations of $\hslck$ do
exist \cite{KW88}.
The character formulae obtained in \cite{BHT97} as functions of three variables $\tau, \sigma$ and $\nu$  are not directly suitable for the analysis of their
behaviour under the modular group. In order to make this analysis straightforward, we provide here a decomposition of the $\hslck$ characters
into characters of the subalgebra $\iso_k$. It turns out that for a level of
the form,
\ben
k=\frac{1}{u}-1,~~~ u=2,3,...,
\label{level1}
\end{equation}
i.e. of the form \eqref{levelp} with $p=1$, the branching functions involve 
characters of a 
rational torus model $A_{u(u-1)}$ as well as $\Z _{u-1}$ parafermion characters. The identification of the branching functions relies on the residue analysis at the 
simple pole developed, in the limit where $\sigma$ tends to zero, by a subset of $\hslck$ characters.
One particular feature of $\super$ is that the central charge of the associated
Virasoro algebra is zero (a consequence of equal number of bosonic and fermionic
generators in $sl(2/1)$),
\ben
c_{sl(2|1)}=0,
\end{equation}
so that the coset $\frac{\super}{\iso _k}$ has central charge 
$c_{\rm coset}=c_{sl(2|1)}-c_{sl(2)}=-\frac{3k}{k+2}$. 

Since we restrict ourselves here
to a family of conjectured rational theories at levels of the form 
\eqref{level1}, 
the coset central charge is positive,
\ben
c_{\rm {coset}}=\frac{3(u-1)}{u+1},
\label{ccc}\end{equation}
and its value is one when $u=2$.  
The branching functions in this case are just the characters
of the rational torus algebra $A_2$, as already noticed in \cite{BHT97}.

The paper is organised as follows. In section 2, the
$\super$ characters given in \cite{BHT97} are rewritten
in terms of infinite products and  theta functions. The latter formulation
leads in section 3 
to the decomposition of $\super$ characters into $\iso _k$ characters.
 Section 4 uses the residues of
singular $\super$ characters at the pole $\sigma =0$ to rewrite the branching
functions in such a way that the S transform of the $\super$ characters can be easily worked out. The decomposition formulae \eqref{string1} and \eqref{string2}
are conjectured for higher values of the parameter $u$ on the basis of a
detailed analysis of the cases $u=3,4$ sketched in the appendix.
We offer some comments on the structure of the coset
$\frac{\super}{\iso _k}$ in our conclusions.

\section{New formulae for $\super$ admissible characters.}
\renewcommand{\theequation}{2.\arabic{equation}}
\setcounter{equation}{0}
In \cite{BHT97}, we derived formulae for the characters of irreducible representations of $\super$ at fractional level $k=\frac{p}{u}-1$,
where $p$ and $u$ are positive coprime integers. Those characters are obtained
in a standard way from the knowledge of the quantum numbers 
and embedding diagrams of singular vectors within a given Verma module
\cite{BT96}. 
By definition, $\super$ characters are given by,
\ben
\chi^{\super}_{h_-,h_+}
(\tau,\sigma,\nu)\defd\text{tr}\exp\bigl(2\pi \mathrm{i}(\tau L_{0}+\sigma
J_{0}^{3}+\nu U_{0})\bigr)
\end{equation}
where $J_0^3$ and $U_0$ are the zero modes of the $\super$ Cartan generators, 
and the isospin $\hf h_-$ and charge $\hf h_+$ are their eigenvalues when they act on the highest weight state $|\Omega \rangle$ of the representation,
\ben
J_0^3 |\Omega \rangle = \hf h_- |\Omega \rangle,~~~U_0 |\Omega \rangle=\hf h_+ |\Omega \rangle.
\end{equation}
The variables $q,~z$ and $\zeta$ are defined by,
\begin{align}
 & q\defd\exp(2\pi \mathrm{i}\tau), \quad \tau \in \mathbb C \quad   
      \im(\tau)>0\Rightarrow|q|<1,\notag\\
 & z\defd\exp(2\pi \mathrm{i}\sigma), \quad \sigma\in\mathbb C,\notag\\
 & \zeta\defd\exp(2\pi \mathrm{i}\nu), \quad \nu\in\mathbb C.
\end{align}

For instance, the class IV Ramond characters read,
\begin{gather}
\chi^{R,IV,\super}_{h_-^R,h_+^R}(\tau,\sigma,\nu)=q^{h^{R}}z^{\frac{1}{2} h_{-}^{R}}
\zeta^{\frac{1}{2}h_{+}^{R}} F^R(\tau,\sigma, \nu)\times \notag\\[3mm]
\sum_{a\in\Z}q^{a^{2}pu+apm}z^{-ap}\frac{1-q^{2ua+m}z^{-1}}{
(1+q^{au+m'}z^{-\frac{1}{2}}\zeta^{-\frac{1}{2}})(1+q^{au+m-m'}z^{-\frac{1}{2}}\zeta^{\frac{1}{2}})},\label{r4char}
\end{gather}
where $ m,m'\in\Zplus , ~0\leqslant m'\leqslant m\leqslant u-1,$ and
\begin{gather}
h_-^R=-m(k+1),\notag\\
h_+^R=(2m'-m)(k+1).
\label{qnr4}
\end{gather}
By substituting $M+M'+2$ for $m$, $M+1$ for $m'$ and $z^{-1}$ for $z$ in the above,
one obtains an expression for the class V Ramond characters, up to an overall
sign. The range of the parameters $M$ and $M'$ in the Ramond sector class V
is $M,M'\in\Zplus ,~ 0\leqslant M+M'\leqslant u-2$, and the quantum numbers
associated to these parameters are,
\begin{gather}
h_-^R=(M+M'+2)(k+1),\notag\\
h_+^R=(M-M')(k+1).
\label{qnR5}
\end{gather}
In both classes, the conformal weight is given by
\ben
h^R=\frac{1}{4(k+1)}((h_-^R)^2-(h_+^R)^2). 
\end{equation}
On the other hand, the class IV (resp. class V) NS characters are easily
obtained from class IV (resp. class V) R characters by spectral flow,
namely,   
\ben
\chi^{NS,\super}_{h_-^{NS},h_+^{NS}}(\tau,\sigma,\nu)=
q^{k/4}z^{ k/2}\chi^{R,\super}_{h_-^R,h_+^R}(\tau,-\sigma-\tau,\nu),
\end{equation}
where
\begin{align}
h^{NS}&=h^{R}-\half h^{R}_{-}+\quarter k,\notag\\
\half h_{-}^{NS}&=-\half h^{R}_{-}+\half k,\notag\\
\half h_{+}^{NS}&=\half h_{+}^{R}.\label{changedqnos}
\end{align}

The above formulae are not suited to the discussion of the modular properties
of characters. Our ultimate goal is to identify how the above 
$\super$ characters branch into $\iso _k$ characters, in order to make their
modular tranforms straightforward to obtain. We restrict ourselves to levels of the form
$k=\frac{1}{u}-1$, and we
mainly concentrate on the Neveu-Schwarz sector of the theory.

We first use standard techniques to produce expressions for the characters in
terms of infinite products. Namely, we analyse the pole structure of the 
series appearing in the NS version of \eqref{r4char} and manufacture an
infinite product with the same singularities and same residues at the poles.
It is not too difficult to show that,
\begin{multline}
\chi^{NS,IV,\super}_{h_-^{NS},h_+^{NS}}(\tau,\sigma,\nu)=
q^{h^{NS}}z^{\frac{h_{-}^{NS}}{2}}\zeta^{\frac{h_{+}^{NS}}{2}}
F^{NS}(\tau,\sigma,\nu)\times\\
\prod_{n=1}^{\infty}~(1-q^{un})^{2}(1-zq^{un-m})(1-z^{-1}q^{u(n-1)+m})
h_{n}^{-1}(m,m';\tau,\sigma,\nu)
\label{ns4productform}
\end{multline}
where,
\begin{multline}
F^{NS}(\tau,\sigma, \nu) = \\[3mm]
\prod^{\infty}_{n=1}\frac{(1+z^{\frac{1}{2}}\zeta^{\frac{1}{2}}q^{n-\hf})
(1+z^{-\frac{1}{2}}\zeta^{\frac{1}{2}}q^{n-\hf})
(1+z^{\frac{1}{2}}\zeta^{-\frac{1}{2}}q^{n-\hf})
(1+z^{-\frac{1}{2}}\zeta^{-\frac{1}{2}}q^{n-\hf})}
{(1-q^{n})^{2}(1-zq^{n})(1-z^{-1}q^{n-1})},
\label{factorNS}
\end{multline}
and,
\begin{multline}
h_{n}(m,m';\tau,\sigma,\nu)=(1+z^{\frac{1}{2}}\zeta^{\frac{1}{2}}
q^{u(n-1)+m-m'+\frac{1}{2}})
(1+z^{\frac{1}{2}}\zeta^{-\frac{1}{2}}q^{u(n-1)+m'+\frac{1}{2}})\\
(1+z^{-\frac{1}{2}}\zeta^{\frac{1}{2}}q^{un-m'-\frac{1}{2}})
(1+z^{-\frac{1}{2}}\zeta^{-\frac{1}{2}}q^{un+m'-m-\frac{1}{2}}).
\end{multline}

In \eqref{ns4productform}, one has,
\ben
h_-^{NS}=k+m(k+1)~~~~{\rm and}~~~
h_+^{NS}=(2m'-m)(k+1).
\end{equation}
An easy way to obtain the class V NS characters is to substitute $M+M'$ for $m$, $M$ for $m'$ and $z \rightarrow z^{-1}$ in the above expression, and flip the overall sign.

Note that, at fixed level $k=\frac{1}{u}-1$, there exist $u^2$ NS characters in classes
IV and V, $u$ of which are regular at $\sigma =0$. The others develop
a simple pole at this value of $\sigma$ and the residue at the pole is given
by,
\begin{gather}
\lim_{\sigma\rightarrow 0}
2\pi i\sigma\,\chi^{NS,\super}_{h_-^{NS},h_+^{NS}}(\tau,\sigma,\nu) = 
\frac{\vartheta_{0,2}(\tau,\half\nu)+\vartheta_{2,2} (\tau,\half\nu)}{\eta^{3}(\tau)}
\chi^{NS,N=2}_{r,s}(\tau,\zeta^{\hf}),\label{nsres}
\end{gather}
where the $N=2$ superconformal characters appear in the infinite product 
form first derived by Matsuo \cite{Mats}. In the above, the $N=2$ central charge
is $c=3(1-\frac{2}{u})$ and $m=r+s-1,m'=r-\half$ in class IV, while
$M=r-\hf, M'=s-\hf$ in class V.

Similar formulae exist for singular Ramond characters,
\begin{gather}
\lim_{\sigma\rightarrow 0}
2\pi i\sigma\,\chi^{R,\super}_{h_-^R,h_+^R}(\tau,\sigma,\nu)= 
\frac{\vartheta_{1,2}(\tau,\half\nu)+\vartheta_{-1,2} (\tau,\half\nu)}{\eta^{3}(\tau)}
\chi^{R+,N=2}_{r,s}(\tau,\zeta^{\hf}).\label{rres}
\end{gather}

Although interesting, the infinite product formula \eqref{ns4productform} has
a denominator whose behaviour under the modular group is nontrivial. 
Using the Jacobi
triple identity repetitively, as well as the standard properties of theta functions, a tedious
calculation leads from \eqref{ns4productform} to the following elegant  expression,
\begin{gather}
\chi_{h_-^{NS},h_+^{NS}}^{NS,IV,\hslck}(\tau,\sigma,\nu)=
\frac{\vartheta_{-u+2(m+1),2u}(\tau,\frac{\sigma}{u})-
            \vartheta_{u+2(m+1),2u}(\tau,\frac{\sigma}{u})}
{\vartheta_{1,2}(\tau,\sigma)-\vartheta_{-1,2}(\tau,\sigma)}
\eta(\tau)^{-1}\eta^{3-2u}(u\tau)\times\notag\\[2mm]
 \prod_{r=1}^{u-1} \sum_{s=0}^{1}
\vartheta_{us+m-2m',u}(\tau, \frac{\nu}{u})
\vartheta_{u(s+1)+m+1+2r,u}(\tau,\frac{\sigma}{u}).
\label{ns4thetaform}\end{gather}

The last obstacle to easy modular transformations is the presence of the function $\eta^{3-2u}(u\tau)$.
Our strategy to eliminate this function consists of two steps. The first  
is to rewrite the expression \eqref{ns4thetaform} in terms of $\iso _k$ characters, 
as described in the next section. The branching coefficients are 
functions of $\nu$ and $\tau$ and still involve the function
$\eta^{3-2u}(u\tau)$. 
The second step, described in section~4, eliminates this function
from the branching coefficients by calculating the residue at the pole 
$\sigma =0$ of each singular $\super$ character when decomposed in $\iso _k$ characters (formulae \eqref{string1},\eqref{string2}), and comparing the result obtained with the expressions
\eqref{nsres} and \eqref{rres}.

\section{Branching $\super$ into $\iso _k$.}
\renewcommand{\theequation}{3.\arabic{equation}}
\setcounter{equation}{0}

 The $\iso_k$ characters 
can be written as, see \cite{KW88,MP}
\ben
\chi^{\iso _k}_{n,n'}(\tau,\sigma)
        =\frac{\vartheta_{b_{+},a}(\tau,\textstyle{\sigma\over u})
           -\vartheta_{b_{-},a}(\tau,\textstyle{\sigma\over u})}                                                                               {\vartheta_{1,2}(\tau, \sigma)-\vartheta_{-1,2}(\tau, \sigma)}, 
    \label{su2ad}
\end{equation}
where the level is parametrized as,
\ben
k=\frac{t}{u},~~~~~{\rm gcd}(t,u)=1,~~~u \in \N,~~t \in \Z,
\label{level}
\end{equation}
with
$ 0\leqslant n\leqslant 2u+t-2 \text{ and } 0\leqslant n'\leqslant u-1$ and,
\ben
b_{\pm}\defd u(\pm(n+1)-n'(k+2)),\qquad a\defd u^{2}(k+2).
\end{equation}
In order to identify which $\iso_k$ characters enter in the decomposition
of the class IV, NS $\hslc$ characters, we
first rewrite \eqref{ns4thetaform} as,
\begin{multline}
\chi_{h_-^{NS},h_+^{NS}}^{NS,IV,\hslck}(\tau,\sigma,\nu)=
\frac{\vartheta_{-u+2(m+1),2u}(\tau,\frac{\sigma}{u})-
            \vartheta_{u+2(m+1),2u}(\tau,\frac{\sigma}{u})}
{\vartheta_{1,2}(\tau,\sigma)-\vartheta_{-1,2}(\tau,\sigma)}
\eta(\tau)^{-1}\eta^{3-2u}(u\tau)\times\\[2mm]
\sum_{n=1}^u ~\biggl\{ \theta_{m-2m',u}(\tau, \frac{\nu}{u})^{u-n}
             \theta_{m-2m'+u,u}(\tau, \frac{\nu}{u})^{n-1}\times\\[2mm]
\sum_{ \{p_i \}_{i=1}^{u-n} \subset S}
\prod_{i=1}^{u-n} \theta_{m+1+u+2p_i,u}(\tau, \frac{\sigma}{u})
\prod_{j=1}^{n-1} \theta_{m+1+u+2p_{u-n+j},u}(\tau, \frac{\sigma}{u})\biggr\},
\label{ns4bin}\end{multline}
where the sum
$
\sum_{\{p_i\}_{i=1}^{u-n} \subset S}
$
is over the $\frac{(u-1)!}{(u-n)!(n-1)!}$ possible subsets $S_{(n)}$ of $(u-n)$ distinct integers $p_i$ included in the set $S=\{1,...,u-1\}$. For each choice
of subset $S_{(n)}$, the variables $p_{u-n+1},...,p_{u-1}$ take the distinct values in $S \backslash S_{(n)}$. 

The following expression
is central in our discussion of branching functions. It reads,
\begin{multline}
\chi_{h_-^{NS},h_+^{NS}}^{NS,IV}(\tau,\sigma,\nu)=
\eta(\tau)^{-1}\eta^{3-2u}(u\tau)\times
\sum_{n=1}^u ~~\Biggl\{ {\cal{F}}(n;\tau, \frac{\nu}{u})\times\\
\sum _{
\{ p_i \}_{i=1}^{u-n} \subset S}  
\biggl\{~~
\sum_{r=0}^{u-2} ~~\Bigl\{
\sum_{ D(\mu _1,...,\mu _{u-n-1};\nu _1,...,\nu _{n-2};r)} \bigl\{ ~~
{\cal{G}}(p_1,...,p_{u-1};\tilde{\mu};\tilde{\nu})\times\\ 
\sum_{\ell=0}^{u-2} ~\{ 
  \theta_{u(u-n)(n-1)(1-2\ell)
          +2(n-1)(\bar{p}_{u-n}-u\tilde{\mu}_1)
          -2(u-n)(\bar{p}_{u-n,n-1}-u\tilde{\nu}_1), u(u-1)(u-n)(n-1)}  (\tau)
\times\\
\sum_{\lambda=0}^u 
  \theta_{u [(u-1)(4\lambda+3)+2(u-n)(1-2\ell)-4r], 2u(u-1)(u+1)}  (\tau) 
  (-1)^{\epsilon} \chi^{\iso}_{[n_{\epsilon}],u-m-1}  (\tau, \sigma)
 \} \bigr\} \Bigr\} \biggr\} \Biggr\}.
\label{keyex}\end{multline}

In the above formula, the quantum numbers of the representation considered
are ,
\ben
h_-^{NS}=-\frac{1}{u}(u-m-1),~~~~~{h_+}^{NS}=\frac{1}{u}(2m'-m),~~~~~0 \le m' \le
m \le u-1.
\label{qn}\end{equation}
Given a set of $n$ integers $\alpha_i, i=1,..,n$, one also introduces,
\ben
\tilde{\alpha}_i = \sum_{j=i}^{n} (n+1-j) \alpha_j
\end{equation}
and the domain,
\begin{multline}
D(\alpha_1,...,\alpha_{n};r)=\\
\bigl\{ \alpha_j ~:~0 \le \alpha_j \le n+1-j,~j=1,\ldots ,n~:
 \tilde{\alpha}_1=k'(n+1)+r, k' \in \N \bigr\},
\end{multline}
with
\ben
0 \le \tilde{\alpha}_1 \le  \frac{1}{6}n(n+1)(2n+1).
\end{equation}
In particular, one has,
\begin{multline}
D(\mu_1,...,\mu_{u-n-1};\nu_1,...,\nu_{n-2};r)=
\bigl\{ (\mu_j;\nu_{j'}) : 
                  0 \le \mu_j \le u-n-j,~j=1,\ldots ,u-n-1;\\
                  0 \le \nu_{j'} \le n-1-j',~j'=1,\ldots ,n-2: \tilde{\mu}_1+\tilde{\nu}_1=k'(u-1)+r, k' \in \N \bigr\},
\end{multline}
with
\begin{align}
&0 \le \tilde{\mu}_1 \le  \frac{1}{6}(u-n-1)(u-n)(2(u-n-1)+1),\notag\\[2mm]
&0 \le \tilde{\nu}_1 \le  \frac{1}{6}(n-2)(n-1)(2(n-2)+1). 
\end{align}

The function ${\cal{G}}$ is given by the following product,
\begin{multline}
{\cal{G}}(p_1,...,p_{u-1};\tilde{\mu};\tilde{\nu})=
\prod_{i=1}^{u-n-1} \theta_{2P(0; u-n-i)-2u\tilde{\mu}_i,(u-n-i)(u-n-i+1)u}
(\tau) \\
\times \prod_{j=1}^{n-2}\theta_{2P(u-n;n-1-j)-2u\tilde{\nu}_j,(n-1-j)(n-j)u}
(\tau),
\end{multline}
where one defines,
\ben
P(\alpha;\beta)=\bar{p}_{\alpha,\beta} -\beta p_{\alpha +\beta +1},
\end{equation}
with
\ben
\bar{p}_{j,n}=\sum_{k=1}^n p_{j+k},~~~\bar{p}_{0,n} \equiv \bar{p}_n.
\end{equation}

The function ${\cal{F}}$ reads,
\begin{multline}
{\cal{F}}(n;\tau,\frac{\nu}{u})=
\sum_{s=0}^{u-n-1} \biggl\{ 
\sum_{t=0}^{n-2} \Bigl\{ 
\sum_{D(\rho _1,...,\rho _{u-n-1};s)} \bigl\{
\sum_{D(\sigma _1,...,\sigma _{n-2};t)} \{ 
{\cal{G}}(0; \tilde{\rho};\tilde{\sigma}) \\[2mm]
\sum_{\lambda '=0}^{u-2} 
   \theta_{u [-(u-n)(n-1)(2\lambda '+1)-2(n-1)s+2(u-n)t],
                                          u(u-1)(u-n)(n-1)}  (\tau)\\  
   \theta_{u [-2(u-n)\lambda'-2s-2t+(n-1)]+(u-1)[m-2m'],      
                                          u(u-1)}   (\tau,\frac{\nu}{u})
\}\bigr\}  \Bigr\} \biggr\}.  \label{F}\end{multline}

Finally, the label $[n_{\epsilon}]$ in the $\iso_k$ characters entering the 
formula \eqref{keyex} is the residue modulo $2(u+1)$ of $n_{\epsilon}$
defined by, ($\epsilon =0,1$)
\ben
n_{\epsilon}=-1~+~\bigl( 1-2\epsilon \bigr)
\bigl(~(u-1)(2\lambda+1)+u-2r+(u-n)(1-2\ell)~\bigr).
\end{equation}
For each choice of variables $\lambda,~ r,~n,~\ell$,
either $[n_0]$ or $[n_1]$ is in the set $S=\{1,...,u-1\}$, or else,
$[n_0]=[n_1]$. In the latter case, there is no contribution proportional
to $\chi^{\iso}_{[n_0],u-m-1}$ in \eqref{keyex}, while in the former case, one gets a contribution $\chi^{\iso}_{[n_{\epsilon}],u-m-1}$ with
$\epsilon =0$ (resp. 1) according to whether $[n_0]$ (resp. $[n_1]$)
is in the set $S$. 

In obtaining \eqref{keyex} from \eqref{ns4bin}, the use of the theta
function identity,
\ben
\theta_{m,k}(\tau, \sigma) \theta_{m',k'}(\tau, \sigma)=
\sum_{\ell=1}^{k+k'} \theta_{mk'-m'k+2\ell kk',kk'(k+k')}(\tau)
\theta_{m+m'+2\ell k,k+k'}(\tau,\sigma)
\end{equation}
is crucial. In particular, it allows to write, 
\begin{equation}
\prod_{i=1}^{n} \theta_{p_i,u}(\tau, \frac{\sigma}{u})=
\sum_{r=0}^{n-1} \bigl\{ \sum_{D(n_1,..,n_{n-1};r)}\prod_{i=1}^{n-1}
\theta_{P(0;n-i)-2u\tilde{n}_i,(n-i)(n-i+1)u}(\tau)\bigr\}
\theta_{\bar{p}_n-2ur,nu}(\tau,\frac{\sigma}{u}).
\label{superfor}\end{equation}
Note that the invariance of the left hand side of \eqref{superfor}
under permutations of $\{ p_1,...,p_n \}$ provides identities between
sums of products of theta functions at fixed values of $r$. This type of identity has been used in deriving \eqref{keyex}.
Let us end this section by noting that the corresponding decomposition
in $\iso_k$ characters for NS class V $\hslc$ characters is readily
obtained by making the substitution $M+M'$ for $m$, $M$ for $m'$, $z \rightarrow z^{-1}$ and an overall flip of sign in \eqref{keyex}. 

\section{Parafermionic characters as branching functions.}
\renewcommand{\theequation}{4.\arabic{equation}}
\setcounter{equation}{0}

As stressed in the introduction, the expression \eqref{keyex} neatly
isolates the 
$\iso_k$ character dependence, but the branching functions
are still written in a way which obscures their modular properties.
However, it is easy to calculate the residue at the simple pole $\sigma =0$
in the NS singular characters, both in class IV and class V, when they are
decomposed in $\iso_k$ characters. Indeed, as discussed in \cite{MP},
when the level $k$ is of the form \eqref{level1}, the residue at the pole $\sigma = 0$ of singular $\iso_k$
characters are unitary minimal Virasoro characters at level $u$,
multiplied by $\eta^{-2}(\tau)$. 

On the other hand, we have the residue calculated as in \eqref{nsres},
which can be rewritten using $\su$ string functions at level $(u-2)$ \cite{KP,RY}, \newpage
\begin{eqnarray}
&&\lim_{\sigma\rightarrow 0}
2\pi i\sigma\,\chi^{NS,\super}_{h_-^{NS},h_+^{NS}}(\tau,\sigma,\nu) = 
\frac{\vartheta_{0,2}(\tau,\half\nu)+\vartheta_{2,2} (\tau,\half\nu)}{\eta^{3}(\tau)}
\chi^{NS,N=2}_{r,s}(\tau,\zeta^{\hf})\cr
&&=\frac{\vartheta_{0,2}(\tau,\half\nu)+\vartheta_{2,2} (\tau,\half\nu)}{\eta^{3}(\tau)}
\sum_{\tilde{m}'=-u+3}^{u-2} c^{(u-2)}_{\tilde{\ell},\tilde{m}'}(\tau)
\theta_{\tilde{m}'u-\tilde{m}(u-2),u(u-2)}(\frac{\tau}{2},\frac{\nu}{2u})\cr
&&=\frac{1}{\eta^3(\tau)}
\sum_{\tilde{m}'=-u+3}^{u-2} c^{(u-2)}_{\tilde{\ell},\tilde{m}'}(\tau)~~
\biggl\{ \sum_{r=0}^{u-2} 
\bigl\{ \sum_{s=0}^1 \theta_{\hf (\tilde{m}'u-\tilde{m}(u-2))-(2r-s)(u-2),
(u-1)(u-2)}(\tau)\cr
&&~~~~~~~~~~~~~~~~~~~~~~~~~~~~~\times \theta_{\hf (\tilde{m}'u-\tilde{m}(u-2))+u(2r-s),u(u-1)}
(\tau,\frac{\nu}{u}) \bigr \} \biggr\},
\label{res}\end{eqnarray}
where 
\ben
\tilde{\ell}=m=r+s-1~~{\rm and}~~\tilde{m}=2m'-m
\end{equation}
in class IV.

The above residues are thus expanded in a basis of theta functions at level
$u(u-1)$ with arguments $\tau$ and $\frac{\nu}{u}$, as are the residues
calculated from \eqref{keyex} (see \eqref{F}). A comparison between these two methods of
obtaining the residues of singular NS $\hslc$ characters provides enough
information to express the branching functions in a form where the modular
transformations can be carried out easily. We have worked out in detail
the branching functions for the cases $u=3$ and $u=4$ (see the appendix), and the truly remarkable result
is that the branching functions involve a rational torus $A_{u(u-1)}$ \cite{DVV}
as well as $\Z _{u-1}$ parafermion characters \cite{ZAM,Qiu}. With hindsight, this structure can be derived from the coset central charge value \eqref{ccc}, since
\ben
c_{\rm {coset}}=\frac{3(u-1)}{u+1}=1~+~\frac{2(u-2)}{u+1},
\label{ccoset}\end{equation}
where the first term is the torus central charge while the second term is precisely the central charge of the parafermionic algebra based on $\Z_{u-1}$ parafermions. When $u=2$, there are no parafermions and the branching functions
are just the $A_2$ torus characters, as discovered in \cite{BHT97}.
Based on the analysis for low values of the parameter $u$, we conjecture
the following $\hslck$ character decomposition formulae. The class
IV characters can be written,
\begin{gather}
\chi_{h^{NS}_-,h^{NS}_+}^{NS,IV,\hslck}(\tau,\sigma, \nu)=
\sum_{i=0}^{u-1}  \chi^{\iso_k}_{i,u-m-1}(\tau,\sigma)\times\notag\\
\left\{\sum _{s=0}^{u-2}
c^{(u-1)}_{i, 3i+2us}(\tau) \theta_{(u-1)(m-2m')+u(u-1)(i+1)+2iu(\frac{u}{2}-[\frac{u}{2}])-2us,u(u-1)}
(\tau, \frac{\nu}{u})\right\},
\label{string1}\end{gather}
where $h_-^{NS}$ and $h_+^{NS}$ are given by \eqref{qn},
while 
the class V NS characters are decomposed as, 
\begin{gather}
\chi_{h^{NS}_-,h^{NS}_+}^{NS,V,\hslck}(\tau,\sigma, \nu)=
\sum_{i=0}^{u-1} \chi^{\iso_k}_{i,M+M'+1}(\tau,\sigma)\times\notag\\
\left\{\sum _{s=0}^{u-2}
c^{(u-1)}_{i, 3i+2us}(\tau) \theta_{(u-1)(M'-M)+u(u-1)i+2iu(\frac{u}{2}-[\frac{u}{2}])-2us,u(u-1)}
(\tau, \frac{\nu}{u})\right\},
\label{string2}\end{gather}
where 
\ben
h_-^{NS}=-\frac{1}{u}(M+M'+1+u),~~~~~{\rm and}~~~~~h_+^{NS}=\frac{1}{u}(M-M').
\label{qn5}\end{equation}
The symbol $[\frac{u}{2}]$ is the integer part of $\frac{u}{2}$.
In the above expressions, one interprets 
$\eta(\tau)c^{(u-1)}_{i, -3i-2us}(\tau)$ as the partition function for the 
$\Z_{u-1}$ parafermionic theory where the lowest dimensional field is
$\Phi^i_{-3i-2us}$ in the notations of \cite{GepQiu}. Recall that the $\su$ string functions at level $u-1$ have the following symmetries \cite{KP},
\begin{gather}
c^{(u-1)}_{\ell,m}(\tau)=c^{(u-1)}_{\ell,-m}(\tau)=
c^{(u-1)}_{\ell,m+2(u-1)\Z}(\tau)=c^{(u-1)}_{u-1-\ell,u-1-m}(\tau),\notag\\
c^{(u-1)}_{\ell,m}(\tau)=0 ~{\rm for}~~\ell-m \neq 0~~{\rm mod}~2.
\end{gather}
The $A_{u(u-1)}$ torus characters are given by the theta functions at level
$u(u-1)$ multiplied by $\eta^{-1}(\tau)$. The decomposition \eqref{string1}
encodes all information needed to obtain a parafermionic realisation of
$\hslck$ at fractional level \cite{bht98}. 

\section{Conclusions}
\renewcommand{\theequation}{5.\arabic{equation}}
\setcounter{equation}{0}

Admissible representations of affine Lie algebras and superalgebras have been 
known to exist for fractional levels since the pioneering work of Kac and Wakimoto \cite{KW88}. It is argued there that the corresponding characters can be expressed in terms of modular forms and provide a finite representation of the modular group.

Over the last two years, we have calculated
explicit character formulae for the affine superalgebra $\hslck$  using
the Kac-Kazhdan determinant formula \cite{dob,KK} and the embedding diagrams for
singular vectors one can derive from it \cite{BT96,BHT97}. A necessary
condition for admissibility is that the level of the algebra
$\hslck$ is of the form \eqref{levelp}. In this paper, we restrict ourselves to levels of the form $k=\frac{1}{u}-1,~u \in \N$, and to a certain class of highest weight
representations labeled class IV and class V in \cite{BHT97}, where the
highest weight state quantum numbers are given by \eqref{qn} and \eqref{qn5}.
The corresponding characters, first derived in \cite{BHT97}, are decomposed
here into characters of the bosonic subalgebra $\iso_k$, which has been extensively studied.  The branching functions involve characters of the rational torus $A_{u(u-1)}$ and
partition functions of the parafermionic algebra $\Z_{u-1}$, given by $\su$ string
functions at level $(u-1)$. The decomposition \eqref{string1} and \eqref{string2}
are conjectured for arbitrary value of $u \in \N$ on the basis of a detailed 
analysis of the cases $u=2$ \cite{BHT97} and $u=3,4$ (see Appendix). They reflect the structure of the central charge of the coset $\hslck/\iso_k$,
given by \eqref{ccoset} as the sum of a torus and a $\Z_{u-1}$ parafermionic
central charge.

There are three immediate byproducts of these decomposition formulae.
First of all, the S  modular transform of the $\hslck$ characters
can easily be obtained from \eqref{string1} and \eqref{string2} since the
modular transformations laws of theta functions, $\su$ string functions and $\iso_k$
characters are standard. We have explicitly checked that in the cases $u=2,3$,
the irreducible $\hslck$ characters provide a finite representation of the modular group, indicating the existence of an 
underlying rational theory.
Second, as pointed out in the appendix, the more technical steps taken
in obtaining the branching functions produce mathematical identities reminiscent
of those discovered and discussed in \cite{Anne,TW97}, and may well put the
latter in a new perspective.
Third, the decomposition formulae encode all necessary information to obtain a new representation of the $\hslck$ currents in terms of a primary parafermionic conformal field, as described in \cite{bht98}. In particular,
in the case $u=3$ (resp. $u=4$), the primary field corresponds to the Ising
model (resp. 3-states Potts model) spin field $\sigma$ of conformal weight
$1/16$ (resp. $1/15$). The deep r$\hat{\rm o}$le of such a representation has yet to be
understood, particularly in connection with our interest in the family of levels
studied here, which stems from the non trivial r$\hat{\rm o}$le $\hslck$ plays in the description of non-critical {\em unitary} minimal
$N=2$ strings, when the matter central charge is given by
$c_{\rm  matter}=3 \left(1-\frac{2}{u}\right)$. The most straightforward
generalisation of this work would be to relax the condition $p=1$ and
investigate the theory of nonunitary minimal $N=2$ strings.

\vskip 2cm

{\large\bf {Acknowledgements}}\\

M. Hayes acknowledges the British EPSRC for a studentship. A.Taormina acknowledges the Leverhulme Trust for a fellowship and thanks the EC for 
support under a Training and Mobility of Researchers Grant No. FMRX-CT-96-0012.

\vskip 3cm

 {\Large\bf{Appendix}}
\renewcommand{\theequation}{A.\arabic{equation}}
\setcounter{equation}{0} 
\\

In this appendix, we look in some detail at the derivation
of branching functions for the cases $u=3$ and $u=4$. Incidentally, the
corresponding parafermionic theories are the Ising model and the 3-states
Potts model, a fact which will be highlighted below by the natural
occurrence of unitary Virasoro characters at level 3 and 5 in connection with
$\su$ string functions at level 2 and 3 respectively.
As explained in section 2, at fixed values of $u$,
there exist $u^2$ characters corresponding to irreducible representations, $u$ of which are regular when the variable $\sigma $ tends to zero. We will view the
latter as characters obtained by spectral flow from the other (Ramond or
Neveu-Schwarz) sector, where they are singular. To fix the ideas, let us start
in the NS sector of the $u=3$ theory, where six characters are singular in the
limit described above. We will study in detail the class IV character  labeled
by $h_-^{NS}=-2/3,~h_+^{NS}=0$,
which is obtained
by spectral flow from the (regular) vacuum character in the Ramond sector,
corresponding to $(m,m')=(0,0)$ in \eqref{qnr4}. 

Using the decomposition formula \eqref{keyex}, one gets,
\begin{gather}
\chi^{IV,NS,\hslc_{-2/3}}_{-2/3,0}(\tau,\sigma,\nu)= \eta^{-1}(\tau)\times
\bigl\{ \chi_{0,2}^{\iso}(\tau,\sigma)
[~C_1(\tau)\theta_{0,6}(\tau,\frac{\nu}{3})
+C_2(\tau)\theta_{6,6}(\tau,\frac{\nu}{3})~]\notag\\
+\chi^{\iso}_{2,2}(\tau,\sigma)
[~C_2(\tau)\theta_{0,6}(\tau,\frac{\nu}{3})
+C_1(\tau)\theta_{6,6}(\tau, \frac{\nu}{3})~]
+\chi^{\iso}_{1,2}(\tau,\sigma)
C_3(\tau) [~\theta_{3,6}(\tau, \frac{\nu}{3})
+\theta_{9,6}(\tau,\frac{\nu}{3})~]\bigr\},
\label{decomp3}\end{gather}
where
\begin{align}
C_1(\tau)&= \eta^{-3}(3\tau)[\theta_{15,36}(\tau)-\theta_{33,36}(\tau)]\theta_{2,6}(\tau)
+[\theta_{3,36}(\tau)-\theta_{21,36}(\tau)]\theta_{4,6}(\tau),\notag\\[2mm]
C_2(\tau)&= \eta^{-3}(3\tau)[\theta_{15,36}(\tau)-\theta_{33,36}(\tau)]\theta_{4,6}(\tau)
+[\theta_{3,36}(\tau)-\theta_{21,36}(\tau)]\theta_{4,6}(\tau),\notag\\[2mm]
C_3(\tau)&=
 \eta^{-3}(3\tau)[~\theta_{1,6}(\tau)+\theta_{5,6}(\tau)~]
[~\theta_{6,36}(\tau)-\theta_{30,36}(\tau)~].
\end{align}

The residue of the above character when
$\sigma \rightarrow 0$ can be calculated in two different ways. From the decomposition \eqref{decomp3}, it is
readily obtained by noticing that
the three $\iso$ characters have a simple pole singularity 
when $\sigma \rightarrow 0$, and that their residue at the pole are Virasoro characters at level $3$ multiplied by the function $\eta^{-2}(\tau)$ \cite{MP}.
On the other hand, formula \eqref{res} gives the residue in terms of the
unique level
one string function $c^{(1)}_{0,0}(\tau) =\eta^{-1}(\tau)$ (see e.g.\cite{KP}), i.e.
\begin{multline}
\lim_{\sigma \rightarrow 0}
2\pi i\sigma\,\chi^{IV,NS,\hslc_{-2/3}}_{-2/3,0}(\tau,\sigma,\nu) = \eta^{-3}(\tau)
c^{(1)}_{0,0}(\tau)\times\\
\Bigl\{
 \theta_{0,2}(\tau)\theta_{0,6}(\tau,\frac{\nu}{3})
+\theta_{2,2}(\tau)\theta_{6,6}(\tau,\frac{\nu}{3})
+\theta_{1,2}(\tau)[\theta_{3,6}(\tau,\frac{\nu}{3})
                      +\theta_{9,6}(\tau,\frac{\nu}{3})]   \Bigr\}.
\label{res3}\end{multline}
Comparison of these two residue calculations yields the following expressions for the coefficients $C_i(\tau)$,
\begin{align}
[\chi^{Vir(3)}_{1,1}(\tau)^2-\chi^{Vir(3)}_{2,1}(\tau)^2]C_1(\tau)&=
c^{(1)}_{0,0}[\chi^{Vir(3)}_{1,1}(\tau)\theta_{2,2}(\tau)
-\chi_{2,1}^{Vir(3)}(\tau)\theta_{0,2}(\tau)],\notag\\[2mm]
[\chi^{Vir(3)}_{1,1}(\tau)^2-\chi^{Vir(3)}_{2,1}(\tau)^2]C_2(\tau)&=
c^{(1)}_{0,0}[\chi^{Vir(3)}_{1,1}(\tau)\theta_{0,2}(\tau)
-\chi_{2,1}^{Vir(3)}(\tau)\theta_{2,2}(\tau)],\notag\\[2mm]
\chi_{2,2}^{Vir(3)}(\tau)[\chi^{Vir(3)}_{1,1}(\tau)+\chi^{Vir(3)}_{2,1}(\tau)]
C_3(\tau)&=
c^{(1)}_{0,0}\theta_{1,2}(\tau)
[\chi^{Vir(3)}_{1,1}(\tau)+\chi^{Vir(3)}_{2,1}(\tau)],
\end{align}
or again, using well-known identities relating the Ising model characters
to square roots of theta functions at level 2 \cite{ginsp},
\ben
C_1(\tau)=\chi_{2,1}^{Vir(3)}(\tau),~~~
C_2(\tau)=\chi_{1,1}^{Vir(3)}(\tau),~~~
C_3(\tau)=\chi_{2,2}^{Vir(3)}(\tau).
\label{coef}\end{equation}
It is also worthwhile noticing that the three unitary Virasoro characters at level
$3$ coincide with the three partition functions of the $\Z_2$ parafermionic
theory whose lowest dimensional fields are $\Phi^2_2,\Phi^2_0$ and $\Phi^1_1$ 
\cite{GepQiu}. The latter is actually a primary conformal field which is
identified as the spin field of the Ising model.
These partitions are given in terms of $\su$ string functions at level 2. Namely,
\ben
\chi_{1,1}^{Vir(3)}(\tau)=\eta(\tau)c^{(2)}_{2,2}(\tau),~~
\chi_{2,1}^{Vir(3)}(\tau)=\eta(\tau)c^{(2)}_{0,2}(\tau),~~
\chi_{2,2}^{Vir(3)}(\tau)=\eta(\tau)c^{(2)}_{1,1}(\tau),
\end{equation}
and therefore, the above expressions can also be viewed as relating $\su$ string functions at level one and two.

The same analysis can be applied to the other five singular NS $\hslc$ characters at level $k=-2/3$, and to the three singular R characters
which flow to the three regular NS characters at the same level. They all involve the three functions \eqref{coef}.
One 
then obtains expressions which can be read off \eqref{string1},\eqref{string2}
for the nine NS $\hslc$ characters at this level. The modular S transform of
these nine characters is encoded in a $9\times9$ matrix which is unitary
and whose fourth power is the identity \cite{Hayes}.

A similar analysis can be performed for the case $u=4$. For instance, using
again the decomposition formula \eqref{keyex}, one obtains,
\begin{multline}
\chi^{IV,NS}_{-3/4,0}(\tau,\sigma,\nu)=\eta^{-1}(\tau)\times\\
\Bigl\{
\chi^{\iso}_{0,3}(\tau,\sigma)[D_1(\tau)\theta_{12,12}(\tau,\frac{\nu}{4})
  +D_2(\tau)\{\theta_{4,12}(\tau,\frac{\nu}{4})+             \theta_{20,12}(\tau,\frac{\nu}{4})\}]\\
+ \chi^{\iso}_{3,3}(\tau,\sigma)[D_1(\tau)\theta_{0,12}(\tau,\frac{\nu}{4})
  +D_2(\tau)\{\theta_{8,12}(\tau,\frac{\nu}{4})+             \theta_{16,12}(\tau,\frac{\nu}{4})\}]\\
+\chi^{\iso}_{1,3}(\tau,\sigma)[D_3(\tau)\theta_{0,12}(\tau,\frac{\nu}{4})
  +D_4(\tau)\{\theta_{8,12}(\tau,\frac{\nu}{4})+             \theta_{16,12}(\tau,\frac{\nu}{4})\}]\\
+\chi^{\iso}_{2,3}(\tau,\sigma)[D_3(\tau)\theta_{12,12}(\tau,\frac{\nu}{4})
  +D_4(\tau)\{\theta_{4,12}(\tau,\frac{\nu}{4})+             \theta_{20,12}(\tau,\frac{\nu}{4})\}]
\Bigr\},
\label{decomp4}\end{multline}
where the four functions $D_i(\tau),i=1,...,4$ are all sums of quintic
expressions in theta functions (two at level 8, two at level 24 and one at level 120) times the function $\eta^{-5}(4\tau)$. The important remark is that
all sixteen NS $\hslc$ characters at level $k=-3/4$ have the same structure
as the one above, with the same four $D_i$ functions. 
Here again, the residues can be calculated in two ways. The decomposition
\eqref{decomp4} yields Virasoro characters at level 4 multiplied by the function
$\eta^{-2}(\tau)$, while formula \eqref{res} expresses the residue in terms
of $\su$ string functions at level two. By comparison of these two calculations, one
can write,
\begin{align}
\chi^{Vir(3)}_{2,2}(\tau)D_1(\tau)&=
c^{(2)}_{2,0}(\tau)[\chi^{Vir(4)}_{3,3}(\tau)\theta_{6,6}(\tau)  
                    -\chi^{Vir(4)}_{3,2}(\tau)\theta_{0,6}(\tau)]\notag\\[2mm]
&~~~~~+ c^{(2)}_{2,2}(\tau)[\chi^{Vir(4)}_{3,3}(\tau)\theta_{0,6}(\tau)  
                    -\chi^{Vir(4)}_{3,2}(\tau)\theta_{6,6}(\tau)],\notag\\[2mm]
\chi^{Vir(3)}_{2,2}(\tau)D_2(\tau)&=
c^{(2)}_{2,0}(\tau)[\chi^{Vir(4)}_{3,3}(\tau)\theta_{2,6}(\tau)  
                    -\chi^{Vir(4)}_{3,2}(\tau)\theta_{4,6}(\tau)]\notag\\[2mm]
&~~~~~+ c^{(2)}_{2,2}(\tau)[\chi^{Vir(4)}_{3,3}(\tau)\theta_{4,6}(\tau)  
                    -\chi^{Vir(4)}_{3,2}(\tau)\theta_{2,6}(\tau)],\notag\\[2mm]
\chi^{Vir(3)}_{2,2}(\tau)D_3(\tau)&=
c^{(2)}_{2,0}(\tau)[\chi^{Vir(4)}_{1,1}(\tau)\theta_{0,6}(\tau)  
                    -\chi^{Vir(4)}_{3,1}(\tau)\theta_{6,6}(\tau)]\notag\\[2mm]
&~~~~~+ c^{(2)}_{2,2}(\tau)[\chi^{Vir(4)}_{1,1}(\tau)\theta_{6,6}(\tau)  
                    -\chi^{Vir(4)}_{3,1}(\tau)\theta_{0,6}(\tau)],\notag\\[2mm]
\chi^{Vir(3)}_{2,2}(\tau)D_4(\tau)&=
c^{(2)}_{2,0}(\tau)[\chi^{Vir(4)}_{1,1}(\tau)\theta_{4,6}(\tau)  
                    -\chi^{Vir(4)}_{3,1}(\tau)\theta_{2,6}(\tau)]\notag\\[2mm]
&~~~~~+ c^{(2)}_{2,2}(\tau)[\chi^{Vir(4)}_{1,1}(\tau)\theta_{2,6}(\tau)  
                    -\chi^{Vir(4)}_{3,2}(\tau)\theta_{4,6}(\tau)],
\label{rel4}\end{align}
where we have used the identity \cite{Anne},
\ben
\chi^{Vir(4)}_{1,1}(\tau)\chi^{Vir(4)}_{3,3}(\tau) -\chi^{Vir(4)}_{3,1}(\tau)\chi^{Vir(4)}_{3,2}(\tau)=\chi^{Vir(3)}_{2,2}(\tau).
\end{equation}
It turns out that the functions $D_i(\tau)$ enter the partition function
of the 3-states Potts model, and can therefore be interpreted as the partition functions for
the parafermionic algebra $\Z_3$. One has,
\begin{align}
&D_1(\tau)=\chi^{Vir(5)}_{1,1}(\tau)+\chi^{Vir(5)}_{4,1}(\tau)=
\eta(\tau)c^{(3)}_{0,0}(\tau),~~~~~~~~
D_2(\tau)=\chi^{Vir(5)}_{4,3}(\tau)=
\eta(\tau)c^{(3)}_{3,1}(\tau),\notag\\[2mm]
&D_3(\tau)=\chi^{Vir(5)}_{2,1}(\tau)+\chi^{Vir(5)}_{3,1}(\tau)=
\eta(\tau)c^{(3)}_{2,0}(\tau),~~~~~~~~
D_4(\tau)=\chi^{Vir(5)}_{3,3}(\tau)=
\eta(\tau)c^{(3)}_{1,1}(\tau).
\label{coef4}\end{align}
The relations \eqref{rel4} can therefore be viewed as relations between 
$\su$ string functions at level two and level three. It should also be stressed  that the relations \eqref{rel4}, with the functions $D_i$ expressed in
terms of Virasoro characters at level five as in \eqref{coef4}, provide new
identities very similar to the ones obtained in \cite{Anne}.

\def\NPB{Nucl.\ Phys.\ B} 
\def\PRD{Phys.\ Rev.\ D} 
\def\PLB{Phys.\ Lett.\ B} 
\def\MPLA{Mod.\ Phys.\ Lett.\ A}
\def\CMP{Commun.\ Math.\ Phys.}  
\def\IJMPA{Int.\ J.\ Mod.\ Phys.\ A}

\end{document}